\documentclass{sf2a-conf2021}
\usepackage{graphicx}
\usepackage{hyperref}
\usepackage[]{natbib}  
\usepackage{epstopdf}

\def\BibTeX{{\rm B\kern-.05em{\sc i\kern-.025em b}\kern-.08em
    T\kern-.1667em\lower.7ex\hbox{E}\kern-.125emX}}
\bibpunct{(}{)}{;}{a}{}{,}  

\usepackage[T1]{fontenc}

\makeatletter
  \def\@cnjctn{ }
\makeatother

\begin{document}

\TitreGlobal{SF2A 2021}


\title{Extragalactic Globular Clusters with Euclid and other wide surveys}

\runningtitle{Extragalactic GCs with Euclid}

\author{A. Lan\c{c}on}\address{Observatoire astronomique de Strasbourg, Universit\'e de Strasbourg, CNRS, UMR 7550, 67000 Strasbourg, France}

\author{S. Larsen}\address{Department of Astrophysics/IMAPP, Radboud University, PO Box 9010, 6500 GL Nijmegen, The Netherlands}

\author{K. Voggel$^1$}

\author{J.-C. Cuillandre}\address{AIM, CEA, CNRS, Universite Paris-Saclay, Universit\'e Paris Diderot, Sorbonne Paris Cit\'e, Observatoire de Paris, PSL University, F-91191 Gif-sur-Yvette, France}

\author{P.-A. Duc$^1$}

\author{W. Chantereau$^1$}

\author{R. Jain$^1$}

\author{R. S\'anchez-Janssen}\address{UK Astronomy Technology Centre, Royal Observatory, Blackford Hill, Edinburgh, EH9 3HJ, UK}

\author{M. Cantiello}\address{INAF Osservatorio Astr. d'Abruzzo, Via Maggini, 64100 Teramo, Italy}

\author{M. Rejkuba}\address{European Southern Observatory, Karl-Schwarzschild-Strasse 2, D-85748 Garching bei Munchen, Germany}

\author{F. Marleau}\address{Institute fur Astro- und Teilchenphysik, Universit\"at Innsbruck, Technikerstrasse 25/8, A-6020 Innsbruck, Austria}

\author{T. Saifollahi}\address{Kapteyn Astronomical Institute, University of Groningen, PO Box 800, NL-9700 AV Groningen, the Netherlands}

\author{C. Conselice}\address{Jodrell Bank Centre for Astrophysics, University of Manchester, Oxford Road, Manchester M13 9PL, UK}

\author{L. Hunt}\address{INAF - Osservatorio Astrofisico di Arcetri, Largo E. Fermi 5, 50125 Firenze, Italy}

\author{A. M. N Ferguson}\address{Institute for Astronomy, University of Edinburgh Royal Observatory, Blackford Hill, Edinburgh EH9 3HJ, UK}

\author{E. Lagadec}\address{Universit\'e C\^ote d'Azur, 
Observatoire de la C\^ote d'Azur, CNRS, Laboratoire Lagrange, 06304 Nice, France}

\author{P. C\^ot\'e}\address{National Research Council of Canada, Herzberg Astronomy and Astrophysics Research Centre, Victoria, BC V9E 2E7 Canada}

\author{on behalf of the Euclid Consortium}

\setcounter{page}{1}


\maketitle


\begin{abstract}
Globular clusters play a role in many areas of astrophysics, ranging
from stellar physics to cosmology. 
New ground-based optical surveys complemented by observations 
from space-based telescopes with unprecedented near-infrared capabilities
will help us solve the puzzles of their formation histories.
In this context, the Wide Survey of the {\em Euclid}\ space mission will provide 
red and 
near-infrared data over about 15\,000 square degrees of the sky. 
Combined with optical photometry from the ground, it
will allow us to construct a global picture of the globular cluster populations 
in both dense and tenuous environments out to tens of megaparsecs.
The homogeneous photometry of these data sets will rejuvenate 
stellar population studies that depend on precise spectral energy distributions.
We provide a brief overview of these perspectives.
\end{abstract}



\section{Introduction}

A refreshing wave of interest is currently pushing
globular cluster science ahead, triggered by a growing body of
stringent empirical constraints: new formation scenarios are needed 
to explain the abundance patterns seen among globular cluster stars; 
galaxy formation scenarios must explain a variety of globular
cluster color distributions without endagering the scaling relations
between cluster numbers and host galaxy properties;
the relationships between globular clusters (hereafter GC),
dwarf galaxy nuclei and ultra-compact galaxies remain to be elucidated;
direct observations of star forming clumps at high redshift
must find their place in the global picture of globular cluster histories. 
At the same time,
GCs remain objects of reference, with comparatively
simple stellar populations that can allow us to test our understanding
of stellar evolution. For this purpose, at least at distances not yet
accessible to detailed spectroscopic observations, 
homogeneous and deep photometry 
across the spectrum of stellar photospheres is a must.

The nearby future will see the launch of several astronomical telescopes
into space, a few of which will focus on the near-infrared 
spectral range and thus complement large ground-based optical surveys that
are rapidly progressing across the globe. The {\em Euclid} space 
mission\footnote{\url{https://sci.esa.int/web/euclid}}
and the {\em James Webb}\ Space Telescope\footnote{\url{https://www.jwst.nasa.gov/}}
are next in line for launch. While 
the second will provide a variety of instruments for pointed observations
of small fields of view, the first will operate in survey mode
with wide-field cameras in the red part of the optical 
spectrum (VIS instrument) and in the $Y,J,H$ bands of the near-infrared
(NISP instrument). The pointed observations of the {\em James Webb}\  Space Telescope,
later followed by those of the {\em Nancy Grace Roman}\ Space Telescope,
will extend the volume in which Local Group clusters can be resolved
into stars and will also transform our view of high-redshift structure 
formation. The {\em Euclid}\ mission has its place in between these extremes,
and will draw a new picture of GC populations at
intermediate distances, out to almost 100 Mpc.

Globular clusters are among the targets of the 
Legacy Science program of the {\em Euclid}\ mission plan \citep{Laureijs_etal11},
and their detection and study is being prepared as a dedicated work package
of the Local Universe science working group within the Euclid Consortium.
As a consequence of the requirements set by the primary science cases
of {\em Euclid}\ (dark matter and cosmology studies via tracers such as weak lensing 
and galaxy clustering; Mellier, this conference), 
{\em Euclid}\ will observe a large part of the 
darkest skies (almost 15\,000 square degrees; \citealt{Scaramella_etal21}), 
with a good spatial
resolution ($\simeq 0.15''$ in VIS), a well-characterized 
point-spread function, 
and the deep uniform red and near-infrared photometry necessary for photometric
redshift measurements. The need for photometric redshifts has led to
unprecedented coordination efforts with ground-based optical surveys,
in both the Southern and Northern hemispheres.

\section{Taking advantage of {\em Euclid}'s survey specifications}

Deep high-resolution observations of extragalactic GC
populations obtained over the years with the {\em Hubble}\ Space Telescope (HST)
have produced high-purity samples of GCs with information on their half-light
radii out to about 30 Mpc (e.g. 
\citealt{Jordan_etal05, Peng_etal06, Villegas_etal10}),
and GC candidate photometry out to
typically 100 Mpc \citep[e.g. in Coma;][]{Harris_etal09, Saifollahi_etal21},
exceptionally much farther \citep{AlamoMartinez_etal13}. 
However, due to the small field of view of HST cameras,
these data sets are not complete but rather restricted to the pointings 
selected by various observers for their specific purposes. 
Deep pointed observations from the ground have complemented the HST
datasets, again for selected objects; notable
examples are the early-type galaxies of the SLUGGS and MATLAS surveys,
respectively  at $D \leq 27\,{\mathrm{Mpc}}$ and $D \leq 42\,{\mathrm{Mpc}}$
\citep{Brodie_etal14,
Duc_etal20}. The widest ground-based surveys
suitable for extensive GC studies have covered areas of order
$10^2$ square degrees and targeted the dense environments of nearby 
galaxy clusters such as Virgo (NGVS; \citealt{Ferrarese_etal12}) 
or Fornax (NGFS; e.g. \citealt{Briceno_etal18}; FDS; e.g.
\citealt{Cantiello_etal20}).
{\em Euclid}\ will cover almost 15\,000 square
degrees of sky, and this will allow us to characterize GCs around 
galaxies of all types in high- and low-density regions, 
as well as to locate GCs far away from
their host (e.g. \citealt{Jang_etal12, Mackey_etal19}) 
or associated with extended halo substructures (e.g. \citealt{Fensch_etal20}). 
In these external regions dynamical
timescales are longer than near galaxy centers, and the GCs are more
direct tracers of galaxy assembly histories.

\begin{figure}[ht!]
 \centering
 \includegraphics[width=0.72\textwidth,trim=0 160 370 148, clip]{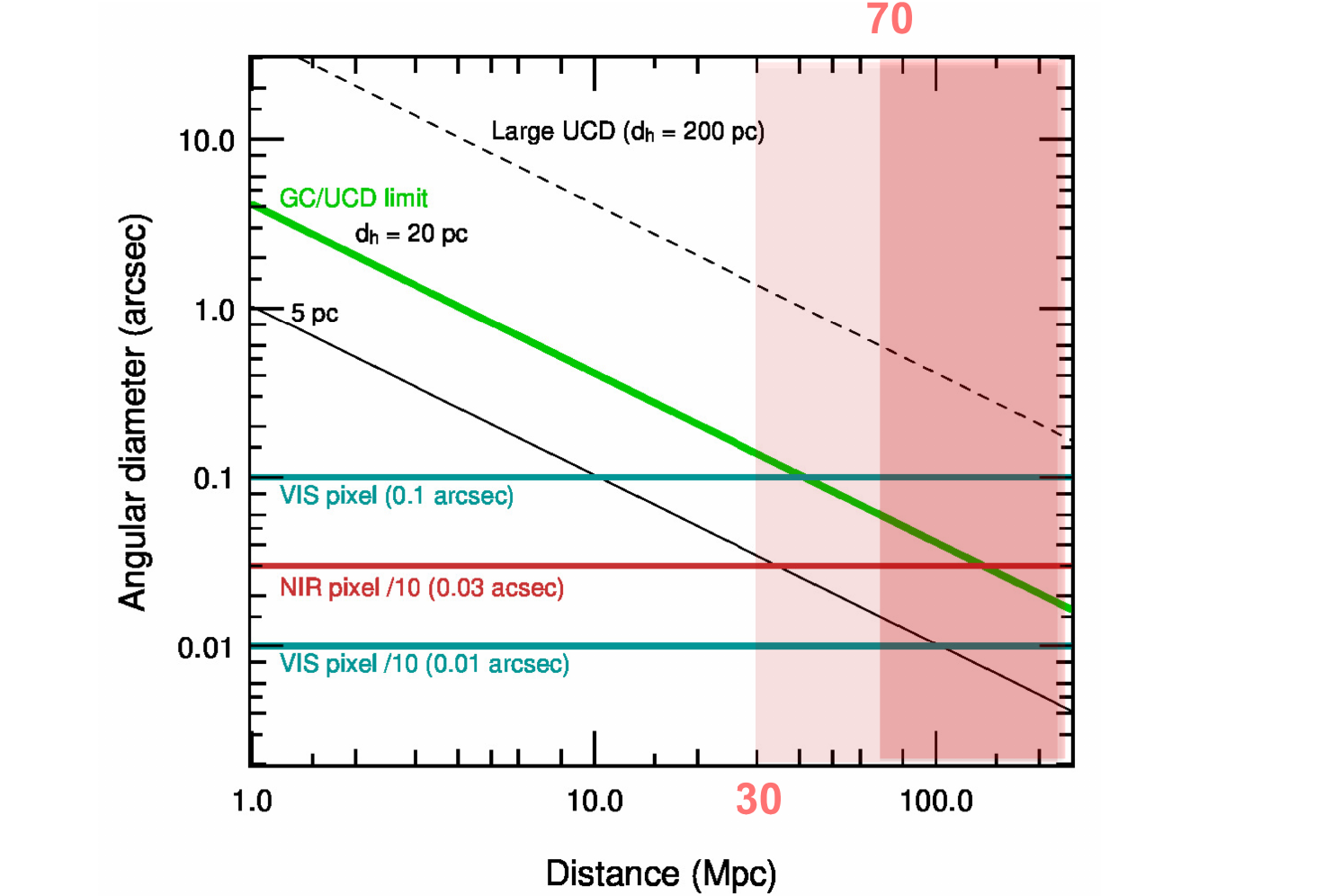}      
  \caption{Properties of globular clusters and ultra-compact dwarf galaxies
  as seen by the {\em Euclid}\ VIS and NIR cameras. The diagonal lines show the
  angular diameters that correspond to half-light diameters
  $d_{\mathrm{h}}$ typical of globular clusters or ultra-compact dwarf galaxies.
  Horizontal lines mark the VIS pixel size, and one tenth of the NIR and 
  VIS pixel sizes. The shaded areas are representative of the limits
  to which GCs near the peak of the GC luminosity function (light shade),
  or bright clusters (dark shade) will be recognized as non point-like.
  }
  \label{ALancon:fig_diam}
\end{figure}

Before GCs can be studied, they must be found. Here, the spatial
resolution of {\em Euclid}'s VIS camera will be an asset. Any catalog
property that reveals the non-point-like nature of a source is
instrumental in separating remote GCs from stellar contaminants,
as was shown in previous studies from the ground (e.g. 
\citealt{Powalka_etal16})
or using a combination of ground-based and Gaia satellite data 
\citep{Voggel_etal20}. The typical 5\,pc 
half-light diameter of a GC will match the VIS
pixel size ($0.1''$) at a distance of 10 Mpc (Fig.\,\ref{ALancon:fig_diam}). 
At a 30 Mpc distance, 
a cluster at the peak of the GC luminosity function 
($M_{\mathrm{VIS}} \simeq -8$\, AB\,mag with some dependence on
color and environment\footnote{Based on \citet{Rejkuba12} and
the $V-\mathrm{VIS}_{\mathrm{AB}}$ indices of stellar population models.})
will have an apparent
AB magnitude of about 24.4, and an expected signal-to-noise
ratio above 20 in stacked VIS images 
(C. Laigle, private communication\footnote{Based on simulations for 
point-sources, Nov. 2020}). 
Its non-point-like nature will be detectable. The non-point-like
nature of brighter clusters will be recognized up to distances
of about 70 Mpc, and this limit will be pushed beyond 100 Mpc
for ultra-compact dwarf galaxies (UCDs).

Within the first few Mpc, the grainy aspect of the semi-resolved outer parts of 
GCs will be a characteristic that the eye and trained
machine-learning algorithms will recognize.
At larger distances however, the separation between GCs and 
redshifted compact objects will require the analysis of colors; the
ideal combination would include {\em Euclid}\ measurements in the red and near-IR
parts of the spectrum, and optical and $u$-band data from other surveys.
Indeed, color-color diagrams that exploit the full photospheric emission 
spectrum of stellar populations are best suited for this exercise
(one now commonly used combination is the $uiK$ diagram; 
\citealt{Munoz_etal14}).
The photometric redshift pipelines that are being developed 
for the core science programs of {\em Euclid}, and that will use data from
ground-based surveys via partnerships, will effectively help rejecting
compact background galaxies. We also intend to implement
dedicated searches that exploit morphology and colors simultaneously.

As already mentioned, {\em Euclid}\ represents a huge step forward in 
near-infrared photometry, not only by pushing the 5\,$\sigma$ detection
limit in $Y,J,H$ to $\sim \,24.4$ AB mag \citep{Scaramella_etal21}
but also by ensuring an excellent uniformity over the sky.  
For the first time, GC spectral energy distributions that include 
near-infrared data will be comparable across samples, 
without the need for color transformations. This will
give new perspectives to studies of the dependencies between the stellar 
populations of GCs and their environment. As both the ground-based
sky surveys of the near future and the Euclid Consortium strive to
improve absolute photometric calibrations, the new data sets will
also provide the most accurate GC colors to date, which will 
allow a critical evaluation of population synthesis model predictions.

\section{How many globular clusters?}

Estimates of the number of GCs that {\em Euclid} catalogs
will contain depend on the final footprint of the wide survey, 
on assumptions on GC specific frequencies and their luminosity distributions, 
and on a number of technical aspects related to the processing
of the images with a pipeline designed primarily for the study
of galaxies in the distant universe.
 
For a first estimate, we have used the NED-D catalog of galaxy 
distances \citep{Steer_etal17},
total optical magnitudes from the Simbad database at 
CDS\footnote{\url{http://simbad.u-strasbg.fr/simbad/} and
\url{http://cdsxmatch.u-strasbg.fr/}}
and a conservative parametrization of the dependence of specific frequencies
on galaxy luminosity based on \citet{Peng_etal08} and 
\citet{Georgiev_etal10}. 
The number of GCs expected to lie within the {\em Euclid}\ footprint out to 35 Mpc
is of several $10^5$, and out to 70 Mpc it exceeds $10^6$. 
How many of these will indeed be measured remains to be examined,
using parameters such as galaxy inclination and pipeline characteristics
such as its ability to measure small sources near large host galaxies.
Rule of thumb estimates suggest that permissive catalogs of GC
candidates will contain a few $10^6$ objects. Excluding the nearest 2\,Mpc
and the largest galaxies in the Virgo and Fornax galaxy clusters (the 
numerous GCs of which have been extensively studied),
the numbers of robust GC candidates with high signal-to-noise 
photometry are expected to approach $10^5$.
Dedicated simulations are being designed to ascertain these 
preliminary values.


\section{Conclusions}

The Euclid Wide Survey will provide us with near-infrared photometry
of unprecedented precision across vast parts of the sky. The combination of
this information with uniform optical photometry from wide ground-based
surveys promises a breakthrough in the study of globular clusters,
in particular in fields that require reliable spectral energy distributions:
the comparison between GC populations in various environments, the 
validation of stellar evolution models and stellar population synthesis 
models for these particular astronomical objects. Indeed, systematic
errors and color transformation uncertainties had become a limiting
factor that the new data will finally push out of the way. 

Complete samples will serve to set up spectroscopic follow-up campaigns, 
leading to new dynamical studies of galaxy halos and of merging histories.
In some areas, {\em Euclid}\ will open doors for deeper studies with pointed 
observations, for instance with the {\em James Webb}\ Space Telescope or the 
{\em Nancy Grace Roman}\ Space Telescope. This will be the case for instance for the 
study of GC populations associated with low surface brightness galaxies
or galaxy sub-structures that Euclid may find, at distances 
of 100 Mpc and more, or for the study of the faint end of 
the GC luminosity function in various environments.

\begin{acknowledgements}
The EC acknowledges the European Space Agency and a number of agencies and institutes that have supported the development of Euclid, in particular the Academy of Finland, the Agenzia Spaziale Italiana, the Belgian Science Policy, the Canadian Euclid Consortium, the Cen- tre National d'Etudes Spatiales, the Deutsches Zentrum f\"ur Luft- und Raumfahrt, the Danish Space Research Institute, the Funda\c{c}ao para a Ci\^encia e Tecnologia, the Ministerio de Economia y Competitividad, the National Aeronautics and Space Administration, the National Astronomical Observatory of Japan, the Netherlandse Onderzoekschool Voor Astronomie, the Norwegian Space Agency, the Romanian Space Agency, the State Secretariat for Education, Research and Innovation (SERI) at the Swiss Space Office (SSO), and the United Kingdom Space Agency (cf http://www.euclid-ec.org). A.L., P.-A.D., W.C., A.L. acknowledge financial support via grant ANR-19-CE31-0022 of the French Agence Nationale de la Recherche.
This work has made use of the SIMBAD database and
of the cross-match service operated at CDS, Strasbourg, France.
\end{acknowledgements}


\bibliographystyle{aa}  
\bibliography{Lancon_S23.bib} 

\end{document}